\documentclass[aps,pra,preprint,superscriptaddress]{revtex4}

\newcommand{\aos}{a_{\mathrm{osc}}}
\newcommand{\rtf}{R_{\mathrm{TF}}}
\newcommand{\beq}{\begin{equation}}
\newcommand{\enq}{\end{equation}}
\newcommand{\rr}{{\bf r}}
\usepackage{graphicx} 
\begin{document}
    
\title{Vortex nucleation in Bose-Einstein condensates in 
time-dependent traps}
\author{Emil Lundh}
\email{Emil.Lundh@helsinki.fi}
\affiliation{Helsinki Institute of Physics, PL 64, FIN-00014
Helsingin yliopisto, Finland}
\author{J.-P. Martikainen}
\affiliation{Helsinki Institute of Physics, PL 64, FIN-00014
Helsingin yliopisto, Finland}
\affiliation{Institute for Theoretical Physics, 
Utrecht University, Leuvenlaan 4, 3584 CE Utrecht, The Netherlands}
\affiliation{Department of Physics, University of Turku, FIN-20014
Turun yliopisto, Finland}
\author{Kalle-Antti Suominen}
\affiliation{Helsinki Institute of Physics, PL 64, FIN-00014
Helsingin yliopisto, Finland}
\affiliation{Department of Physics, University of Turku, FIN-20014
Turun yliopisto, Finland}
\date{\today}

\begin{abstract}
Vortex nucleation in a Bose-Einstein condensate subject to a stirring 
potential is studied numerically using the zero-temperature, 
two-dimensional Gross-Pitaevskii equation. 
In the case of a rotating, slightly anisotropic harmonic potential, 
the numerical results reproduce experimental findings, 
thereby showing that finite temperatures are not necessary for 
vortex excitation below the quadrupole frequency.
In the case of a condensate 
subject to stirring by a narrow rotating potential, the 
process of vortex excitation is described by a classical model that 
treats the multitude of vortices created by the stirrer as a continuously 
distributed vorticity at the center of the cloud, but retains a 
potential flow pattern at large distances from the center.

\end{abstract}
\maketitle

\section{Introduction}\label{sec:introduction}

Since the first experimental creation of a quantized vortex in a 
gaseous Bose-Einstein condensate (BEC) 
\cite{jilavortices,Madison2000a}, there has been
significant experimental and theoretical advance in this field of 
research. However, even in these relatively simple systems, the 
understanding of such a basic issue as the mechanism for nucleation 
of vortices has not been straightforward.

Several experimental methods of vortex creation are currently in use, 
including phase imprinting \cite{jilavortices,Williams1999c},
cooling of a rotating 
normal gas \cite{Haljan2001a}, and conversion of spin angular 
momentum into orbital angular momentum by reversal of the magnetic 
bias field in a Ioffe-Pritchard trap 
\cite{Leanhardt2002a,Nakahara2000a,Leanhardtcoreless}. The topic of 
this paper is the mechanical stirring of a Bose-Einstein 
condensed cloud with the use of optical and magnetic fields. This 
method, in turn, falls into two categories: rotation of an 
anisotropic trap and stirring with a narrow potential, and we shall 
examine each of those in turn.

    References \cite{Madison2000a,foot} reported on
    vortex creation in gases that are confined in 
    traps whose equipotential curves in the $x$-$y$ plane 
    have a slightly elliptic shape 
    and are varied in time so that the axis of ellipticity 
    rotates around the third axis.
    The formation of vortices in such traps is understood to be a 
    consequence of the excitation of unstable quadrupole modes 
    \cite{recati,sinha}. Accordingly, vortices are 
    seen to occur first when the angular frequency of 
    rotation is in the vicinity of $1/\sqrt{2}$ times the average 
    radial trapping frequency $\omega$ (that is, the frequency of the
    quadrupole mode divided by its angular-momentum quantum number)
				\cite{foot,dalibard2}. However, the range 
    of parameter values within which vortices are experimentally seen 
    to be created is not yet entirely accounted for by theory. This shall 
    be discussed in detail later.
    
    Finally, vortices can be excited by the use of a repulsive optical 
    potential which is confined to a small region of the plane, thus 
    mimicking a stick which is moved through the gas cloud in order 
    to stir up vortices \cite{Raman2001a}. In this kind of geometry,
    local 
    turbulence including the creation of vortex-antivortex pairs 
    at the position of the stirrer is instrumental in inducing 
    angular momentum in the gas \cite{Jackson1998}.
    
    In this paper, we set out to investigate to what extent the 
    experimental findings can be accounted for by
two-dimensional zero-temperature theory. We shall do so with means of 
numerical calculations and, when possible, classical hydrodynamical 
modelling.
    The paper is organized as follows. The equation of motion, the 
numerical methods and the elements of the theoretical understanding of 
vortex nucleation are introduced in Section \ref{sec:model}. 
    In Sec.\ \ref{sec:ellipse} 
    we study the formation of vortices in a rotating elliptical trap 
    and compare our numerics with the experimental findings. Section
    \ref{sec:stir} is concerned with the method of stirring the 
    gas with a moving narrow optical potential. In Sec.\ 
    \ref{sec:conclusions} we summarize and conclude.

    \section{Equations and numerical method}\label{sec:model}
    
    We assume the gas cloud to be at sufficiently low temperature and 
    low density 
    that it can be described as a pure condensate. Such a 
    system is entirely governed by a scalar condensate wave function 
    $\psi(\rr,t)$ that evolves in time according to 
    the Gross-Pitaevskii 
    equation \cite{gross,pitaevskii}:
    \begin{equation}
        i\hbar\frac{\partial \psi}{\partial t} =
	-\frac{\hbar^{2}}{2m}\nabla^{2}\psi +
	V(\rr,t)\psi + 
	\frac{4\pi\hbar^{2}a}{m} |\psi|^{2}\psi.
        \label{gpe}
    \end{equation}
The local fluid density and velocity are given in terms of the
condensate wave function by $n(\rr,t) = |\psi(\rr,t)|^2$ and 
$\mathbf{v}(\rr,t) = (\hbar/m) \nabla (\arg \psi(\rr,t))$.
    The nonlinear term in Eq.\ (\ref{gpe}) describes inter-particle
interactions in the 
    s-wave approximation and its strength is determined by the s-wave 
    scattering length $a$.
    $V(\rr,t)$ represents the external potential and can in experiments be
of 
    both magnetic and optical origin. We shall in the following take 
    $V(\rr,t)$ to be the sum of two terms, a two-dimensional
cylindrically 
    symmetric harmonic-oscillator potential and an 
    asymmetric time-dependent term: 
    \beq
    V(x,y,t) = \frac12 m\omega^{2}(x^{2}+y^{2}) + \Delta V(x,y,t).
    \enq
    The spatial extent of the one-particle ground state of the harmonic
potential is called the oscillator length, $\aos=(\hbar/m\omega)^{1/2}$.
    Finally, the condensate wave function $\psi$ is in two dimensions 
    normalized to the number of particles per unit length which we denote 
    $N/l_{z}$, so that $l_z$ is the average extent of the cloud in 
    the $z$ direction. 

    We shall make frequent use of the Thomas-Fermi 
    approximation, which neglects the kinetic term in the 
    Gross-Pitaevskii equation \cite{bp}. This approximation gives an 
    accurate description of the bulk properties of the cloud when the 
    coupling is strong; however, it fails to correctly
    describe short-range density variations such as that at the cloud 
    boundary and close to a vortex core. The stationary-state density 
    in the Thomas-Fermi approximation, assuming an isotropic 
    harmonic-oscillator potential, is
    \begin{equation}\label{tfprofile}
        n(r) = n_{0}\left(1-\frac{r^{2}}{\rtf^{2}}\right)
    \end{equation}
    in the region where the right-hand side is positive and zero 
    elsewhere. The central density is given by
$n_{0} = \mu m/(4\pi\hbar^2a)$; the chemical 
    potential $\mu$ is 
    determined by normalization. In the two-dimensional case 
    considered here, $\mu=2\hbar\omega\sqrt{Na/l_{z}}$. The 
Thomas-Fermi cloud 
    radius is 
    $\rtf=2\,(Na/l_{z})^{1/4}\aos$. Of importance is also the healing 
    length $\xi$, which determines the size of a vortex core; in the 
    Thomas-Fermi approximation it is given by
    $\xi=\left(\aos/\rtf\right)\aos$. The condition of strong coupling 
    is fulfilled when $\rtf \gg \xi$ or, equivalently, when the coupling 
    parameter $Na/l_z$ is large.
    
    In the numerical study, the Gross-Pitaevskii equation
    (\ref{gpe}) is propagated in real time in the nonrotating 
    (laboratory) reference frame, subject to an explicitly 
    time-dependent potential, using the split-step Fourier method. 
	We have chosen not to use any phenomenological 
    model for finite-temperature dissipation such as an imaginary component 
    in the time step 
(cf.\ \cite{Penckwitt2002,Kasamatsu2002,lobo}). 

    As already mentioned above, we have chosen to study only 
    two-dimensional systems, knowing 
    that the physics of vortex formation for oblate systems 
    will be well described by such a model. In fact, as we shall see, 
    the model seems to be able to describe with limited accuracy the 
    essential physics of vortex formation also for a prolate system. 
    
    During the temporal evolution, we have, in addition to 
    the complex wave function itself, monitored the total 
    angular momentum, the mean extension of the cloud along suitably 
    chosen axes (corotating with the disturbance) and the total 
    fluid circulation within a region close to the condensate center. 
    All these quantities are 
    measurable experimentally, and we shall see that each of 
    them exhibits a clear signal in its time evolution when 
    vortices enter the interior of the cloud.
    
For a vortex to be created, it must be energetically favorable 
compared with a nonrotating state \cite{lps}, but in addition, 
there must be a dynamical instability of the cloud that permits 
vortex creation. Generally, an excitation can only be created due to 
a moving disturbance in the fluid when the velocity exceeds a critical
one, given by the Landau criterion \cite{ll}: 
\beq
    v_c = \min_q \frac{E_q}{q},
\enq
where $q$ denotes the wave numbers of all possible quasiparticle 
modes. For a trapped BEC subject to a rotating perturbation 
close to the condensate edge, the Landau analysis yields a critical angular 
velocity in terms of the surface modes \cite{recati,Feder1999}:
\beq\label{landau}
\Omega_c = \min_l\, \frac{\omega_l}{l},
\enq
where $l$ are the angular momentum quantum numbers of the surface 
modes. The result for 
typical trap parameters is a critical linear velocity 
$v_c$ slightly smaller than the speed of sound in the center of 
the trap $c$ \cite{Anglin2001}. As we shall see, this simple analysis 
is not sufficient for the two kinds of geometry studied in the 
present paper, but needs to be modified in different ways.

    \section{Vortex creation in a rotating anisotropic potential}
    \label{sec:ellipse}
    Refs.\ \cite{Madison2000a,foot} describe experiments performed on a
Bose-Einstein condensed cloud contained in a 
harmonic-oscillator potential that has a small 
ellipticity $\epsilon$ in the $x$-$y$ plane which rotates 
with an angular frequency $\Omega$. The time dependent potential 
can be written
    \begin{equation}
        V(x,y,t) = \frac12 m \omega^2\left[(1+\epsilon)X(t)^{2}+
	(1-\epsilon)Y(t)^{2}\right],
    \end{equation}
    where the coordinates along the corotating axes vary in time as
    \begin{equation}
        X(t) = x \cos\Omega t + y \sin \Omega t, \,
        Y(t) = y \cos\Omega t - x \sin \Omega t.
    \end{equation}
The twofold symmetric potential can only excite modes of quadrupolar 
symmetry. Therefore the Landau criterion, Eq.\ (\ref{landau}), must 
be modified so that the summation is restricted to quadrupolar modes, 
resulting in the criterion
    \beq
    \Omega_c = \frac{\omega_2}{2},
    \enq
    where $\omega_2$ is the excitation  frequency of the lowest-lying 
    quadrupole mode. In the Thomas-Fermi limit it is known to have 
    the value $\omega_2=\sqrt{2}\omega$ and therefore the 
    Landau-criterion critical frequency for this trap geometry is
    $\Omega_c = \omega/\sqrt{2} \approx 0.71\, \omega$. 
    However, in order for vortices to be formed, a necessary condition 
    is not 
    only that the quadrupole mode is excited, but also that it 
    is unstable \cite{recati,sinha}. This has been shown to occur when the 
    ellipticity is larger than the critical value
    \begin{equation}
	\epsilon_{0} = \frac{2\omega}{\Omega}\left(
	\frac{2(\Omega/\omega)^{2}-1}{3} \right)^{2/3}.
    \end{equation}

The onset of instability above the quadrupole frequency has been 
investigated in detail in Ref.\ \cite{dalibard2}.
    In Ref.\ \cite{foot}, it was found 
    experimentally that vortices were indeed created only for 
    $\epsilon > \epsilon_{0}$ when $\Omega > \omega/\sqrt{2}$. On the 
    other hand, for driving frequencies slower than the quadrupole 
    frequency, vortices could still be created if the ellipticity 
    exceeded a $\Omega$-dependent value that appears to be well 
    approximated by the straight line 
    $
    \epsilon = 0.71 - \Omega/\omega
    $.
    This can be thought of as a finite width of the resonance at 
    $\Omega = \omega/\sqrt{2}$. In Ref.\ \cite{foot} it was 
    speculated that this finite width is either due to finite 
    temperature or the excitation of surface modes with higher 
    multipolarity $m>2$. The former proposition seems unlikely 
    because there was little temperature variation of the 
    position of the lower threshold value. In addition, 
the experimentally found sub-resonance vortex nucleation is 
reproduced by the zero-temperature theory of Ref.\ \cite{kraemer}, 
as well as by the present numerics, as we shall see.
Reference \cite{kraemer} puts up a pictorial model of vortex entry that
focuses on the path taken by a 
vortex moving into a BEC cloud. 
The finite energy barrier that a vortex would have to overcome in 
order to move into a cylindrically symmetric cloud, 
even under external rotation, is found to be lowered if the cloud is 
elongated. The critical frequency $\Omega$ is determined by the 
requirement that the energy at the saddle point which the vortex has 
to overcome is lower than the energy of a cylindrical, nonrotating 
cloud. The analysis thus applies to the case of a sudden switch-on of 
the rotation. The critical frequency is shifted down from 
the quadrupole frequency by an amount that depends on both $\epsilon$ 
and the coupling strength. However, in the case of an adiabatically 
slow increase 
of the rotational force, the model of Ref.\ \cite{kraemer} would predict 
no vortex entry, because there will always be a local energy minimum 
for a vortex-free elongated state and therefore a finite energy barrier 
for a vortex to overcome.

    The numerical procedure mimics the experimental conditions 
    of Ref.\ \cite{foot}, which has an effective two-dimensional coupling 
    strength $Na/l_z=56$. The two-dimensional calculation neglects any 
    degrees of freedom associated with the $z$ direction, which is a 
    valid approximation since the experimental trap is oblate with an 
    aspect ratio 
    $\omega/\omega_{z} \approx 0.35$ and in such a trap vortices are 
expected to be straight~\cite{GarciaRipoll2001a,Aftalion2001a}. 
The cloud is initially taken to be in 
equilibrium in a stationary trap with zero ellipticity in the 
$x$-$y$ plane. The ellipticity $\epsilon$ is ramped up from 
    zero to its final value over a time $t_{\mathrm{init}}=80\, \omega^{-1}$ 
    while the rotation frequency $\Omega$ is held constant. 
    The subsequent evolution is monitored for another $300\,\omega^{-1}$; 
    typically the vortex nucleation starts after a time comparable to 
    the ramping-up time when $\Omega$ lies above the quadrupole 
    frequency. To obtain a diagnostics for the presence of vortices,
the ellipticity is subsequently ramped down between $t=300\,\omega^{-1}$ 
and $t=380\,\omega^{-1}$, whereafter the vorticity
in an area enclosed by a circle of radius $\rtf$ is computed and averaged 
over a time span of $100\,\omega^{-1}$ in the cylindrically symmetric 
trap. An average vorticity above 0.3 is 
taken to be indicative of the fact that vortices spend a nonnegligible
time in the interior 
of the cloud; the results are not sensitive to the precise threshold value.

\begin{figure}
\includegraphics[width=\columnwidth]{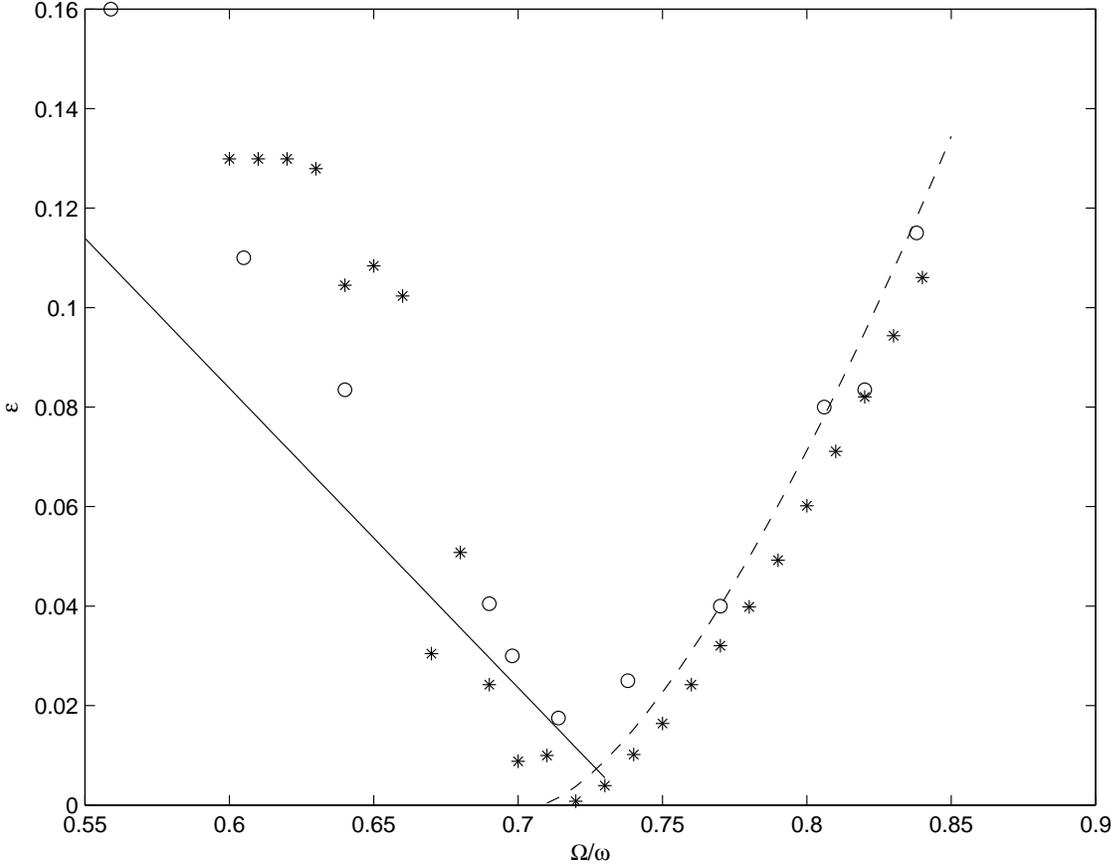}
\caption[fig1]{Critical values of the anisotropy $\epsilon$ of
a rotating anisotropic trap for the onset of vortex nucleation. 
Asterisks denote the numerical results from solving the Gross-Pitaevskii 
equation, the dashed line is the analytical prediction of Ref.\
\cite{recati}, and the circles represent the experimental 
results of Ref.\ \cite{foot}. The solid line represents the analytical 
estimate of Ref.\ \cite{kraemer}, which is a theory for sudden 
switch-on of the rotation, applied to the case of a chemical 
potential $\mu=12\hbar\omega$.
\label{fig:phasediag}}
\end{figure}

    Figure \ref{fig:phasediag} shows the critical value of $\epsilon$ 
    as a function of 
    $\Omega$. 
It is seen that the numerical results 
are in good agreement with the experimental findings and the 
theory of Ref.\ \cite{recati} when $\Omega$ is larger than the 
quadrupole frequency, but the agreement is somewhat poorer below. The 
qualitative features are, however, well captured by our numerics. 
This fact rules out the possibility that vortex excitation 
    below the quadrupole frequency is purely a finite-temperature effect, 
    since the present study is performed using zero-temperature theory.

The prediction by Ref.\ \cite{kraemer} is also in rough agreement 
with the experiment and the present numerics but predicts a lower 
critical anisotropy, which is expected since it applies 
to the case of a sudden switch-on of the trap. We speculate that the 
present ramping-up time, although long compared to the period of the 
forcing $\Omega^{-1}$, is still not perfectly adiabatic.
While the threshold values of $\epsilon$ and $\Omega$ agree with the 
experimental findings of Ref.\ \cite{foot}, the observed numbers of 
vortices turn out to be somewhat higher. This can be due to lack of 
equilibration or because our numerical counting method includes 
vortices on the edge of the condensate that are not visible to the 
eye.

\begin{figure}
\includegraphics[width=\columnwidth]{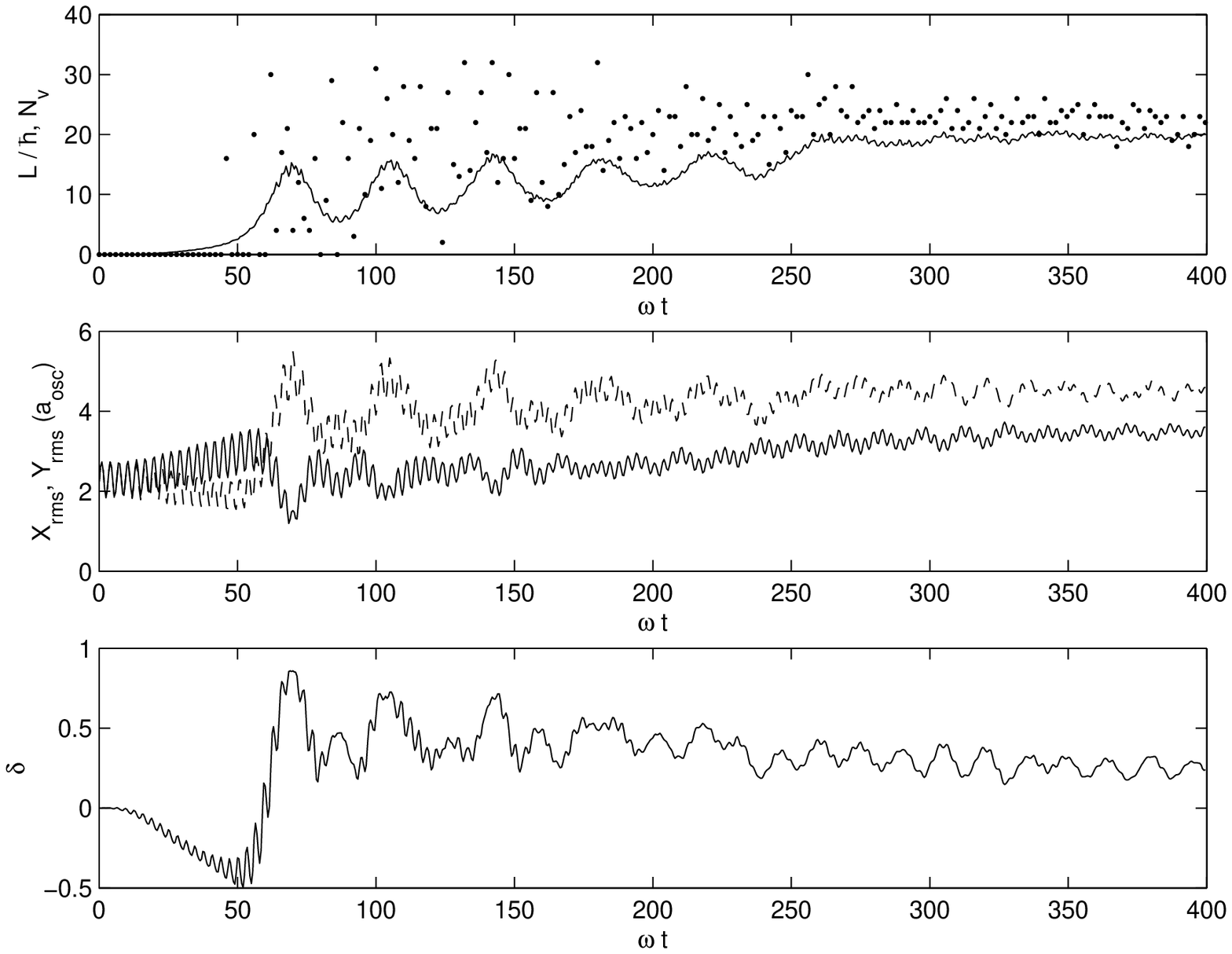}
\includegraphics[width=\columnwidth]{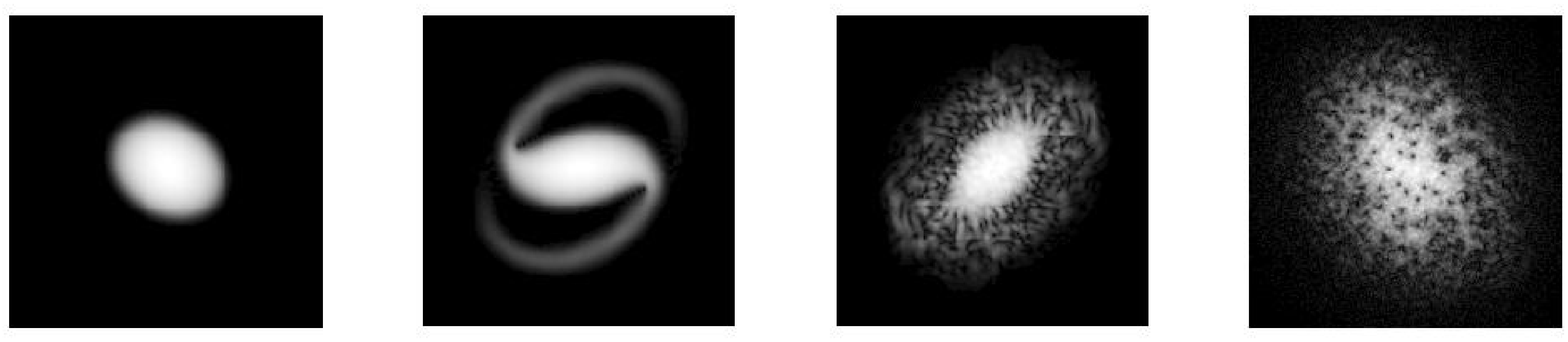}
\caption[]{The upper panel shows the time evolution of the angular 
momentum (full curve) and number of vortices $N_v$ (dots) in
a trap with an anisotropy $\epsilon=0.1$ rotated at a frequency 
$\Omega=0.78\,\omega$. 
The second panel from the top is a plot of the rms radii 
of the cloud along two axes corotating with the anisotropic trap: 
the full curve describes $X_{\mathrm{rms}}=\sqrt{\langle X^2\rangle}$ 
and the dashed curve describes 
$Y_{\mathrm{rms}}=\sqrt{\langle Y^2\rangle}$.
The third panel depicts the dimensionless cloud deformation 
$\delta$ as a function 
of time.
The four pictures on the bottom are density plots taken at the time 
instances $t=27\omega^{-1}$, $t=80\omega^{-1}$, $t=160\omega^{-1}$, 
and $t=400\omega^{-1}$, respectively. Brighter shades represent higher 
density.
\label{fig:lxy}}
\end{figure}

In order to better understand the transition to the vortex state, we 
study in detail the evolution of the $z$ component of the angular 
momentum $L_{z}$ and the mean extension of the cloud along two 
axes corotating with the trap, 
$X_{\mathrm{rms}}=\sqrt{\langle X^{2}\rangle}$ and
$Y_{\mathrm{rms}}=\sqrt{\langle Y^{2}\rangle}$. The deformation of 
the cloud is denoted $\delta$ and defined as 
\beq
\delta = \frac{Y_{\mathrm{rms}}^2-X_{\mathrm{rms}}^2}{Y_{\mathrm{rms}}^2 
+ X_{\mathrm{rms}}^2},
\enq
in accordance with Ref.\ \cite{kraemer}.
In Fig.\ \ref{fig:lxy}, 
these four 
quantities, together with the vorticity $N_v$ in an area enclosed 
by a circle of radius $\rtf$ 
are plotted against time for the choice of parameters 
$\Omega = 0.78 \omega$ and $\epsilon=0.1$. 
    This point lies in the 
    right part of the phase diagram of Fig.\ \ref{fig:phasediag}, 
    above the quadrupole 
    frequency; we shall soon comment on the difference in behavior 
    for different $\Omega$. In this plot, the anisotropy $\epsilon$ is
    initially ramped up for a time $t=80\,\omega^{-1}$ as before, but
    is then held constant throughout the simulation in order to display
    the physics more clearly. We see how the cloud is 
    initially elongated 
    and at the same time rotates with the trap. The cloud 
    initially rotates 90 degrees out of phase with the potential, so that 
    the elongation is in the direction of stronger trapping potential, 
    consistent with Ref.\ \cite{foot}. The 
    elongation of the cloud would now begin to 
    oscillate around an equilibrium value if the frequency and 
    ellipticity were outside the critical region. Within the critical 
    region, the maximum elongation instead 
    grows larger until the outer edges of the cloud bend, close in on 
    themselves, and capture vortices; the first vortex is seen to be 
detected just prior to $t=50\omega^{-1}$ when the cloud elongation is 
at its largest.  After a turbulent period, the 
    anisotropy of the cloud is somewhat decreased. During and after 
    vortex capture, the orientation of the cloud is changed so that the 
    elongation is in the direction of weaker 
    trapping. The whole process is in accordance with 
    previously reported results \cite{Madison2000a,foot,recati,sinha}. 
We do not observe an actual crystallization of the vortices into 
a lattice during the waiting time allowed for here. 
A recent preprint suggests 
that the crystallization of a lattice should occur even at zero 
temperature \cite{lobo}, but the difference in dimensionality, trap 
geometry and number of vortices prevents a quantitative comparison of the 
time scales. 
The effect of dissipative mechanisms on the lattice formation was studied 
in Refs.\ \cite{Penckwitt2002,Kasamatsu2002}.

\begin{figure}
\includegraphics[width=\columnwidth]{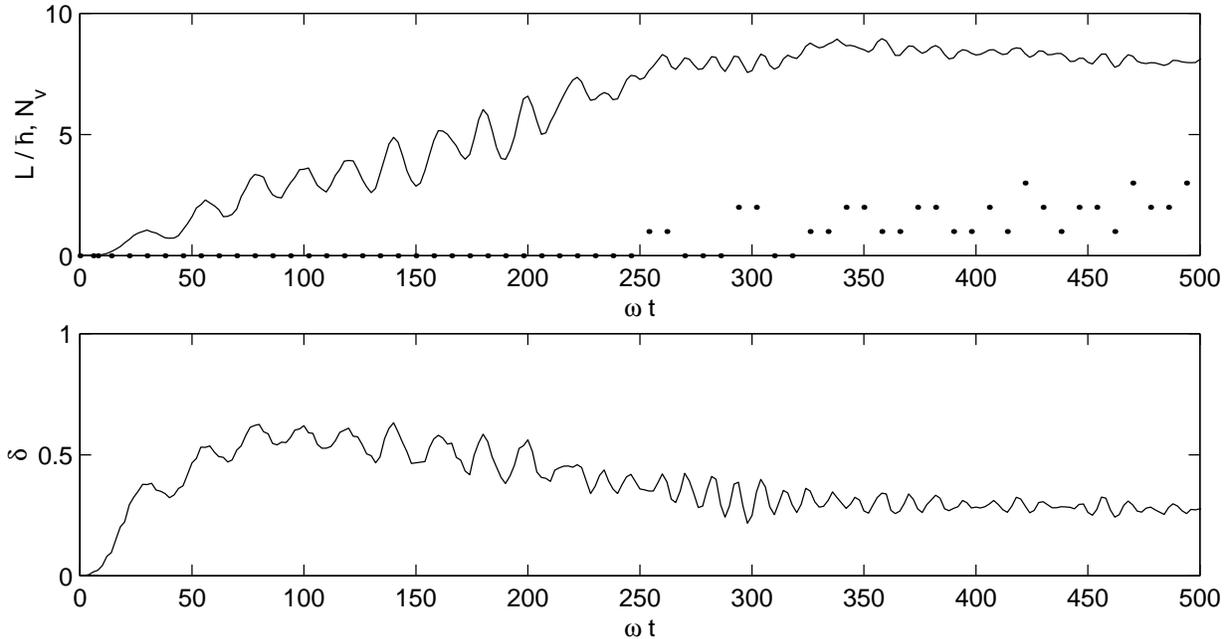}
\caption[]{The upper panel shows the time evolution of the angular 
momentum (full curve) and number of vortices $N_v$ (dots) in
a trap with an anisotropy $\epsilon=0.2$ rotated at a frequency 
$\Omega=0.64\,\omega$. 
The lower panel depicts the dimensionless cloud deformation 
$\delta$ as a function 
of time.
\label{fig:lbelow}}
\end{figure}

Below the quadrupole frequency, the process is markedly different. 
Fig.\ \ref{fig:lbelow} depicts the evolution of the angular 
momentum, vorticity, and cloud deformation $\delta$ in this regime. 
The cloud is initially elongated, but the elongation of the cloud 
rotates in phase 
with the trap deformation so that $\delta$ is positive. The cloud 
deformation is seen to grow to a maximum whereafter vortices enter 
the cloud, consistent 
with the physical picture of Ref.\ \cite{kraemer}. The buildup of 
angular momentum takes place gradually and smoothly compared with the 
situation in Fig.\ \ref{fig:lxy}.

\begin{figure}
\includegraphics[width=\columnwidth]{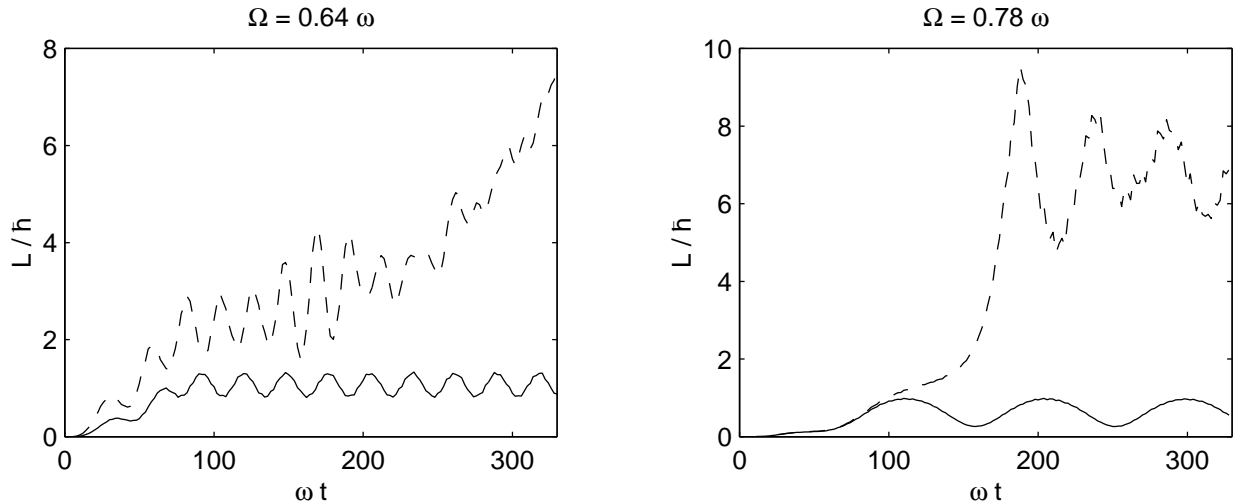}
\caption[]{Time evolution of the angular momentum in a rotating 
anisotropic trap for different values of the rotation frequency
$\Omega$ and trap anisotropy $\epsilon$. In the left panel, the 
rotation frequency $\Omega=0.64\omega$ and the trap deformation 
$\epsilon=0.1$ (full line) and $\epsilon=0.17$ (dashed line). 
In the right panel, the 
rotation frequency $\Omega=0.78\omega$ and the trap deformation 
$\epsilon=0.41$ (full line) and $\epsilon=0.42$ (dashed line).
For the solid lines, 
the parameter values are such that no vortices are created, while the 
oscillations in the case of the dashed lines are associated with vortex 
creation.
\label{fig:fourl}}
\end{figure}

    Figure \ref{fig:fourl} displays the time evolution of the 
    angular momentum for two different choices of $\Omega$, providing 
    representative examples of 
    the evolution of angular momentum below and above the critical 
    line. 
    Clearly, the physics of the 
    vortex creation is different depending on whether the 
    rotation frequency is larger or smaller than the quadrupole 
    frequency. Above the quadrupole frequency (rightmost panel in 
    Fig.\ \ref{fig:fourl}), the vortex formation is 
    very fast, and the oscillations in the vortex-free regime are 
    distinct from the turbulent behavior leading to vortex formation, 
    although the values of $\epsilon$ for the dashed and full curves 
    differ by a very small amount. 
    Below the quadrupole frequency, however, vortex formation is 
    preceded by a gradual buildup of angular momentum, and the outgrowth 
    of ``arms'' from the cloud seems to be absent. In this regime, 
    it takes a 
    finite time for the instability associated with vortex formation 
    to occur. 
    
    The critical lines can thus be determined with good accuracy 
    above the quadrupole frequency, but below the quadrupole 
    frequency, they have a more pronounced dependence on the waiting 
    time. This fact together with the turbulent fluid flow and 
    sensitivity to details that is connected with the vortex creation 
    is the reason for the scatter of the numerical data points in 
    Fig.\ \ref{fig:phasediag}. 
    The gradual buildup of vorticity is consistent with the view that 
    vortices enter the cloud only when there is an appreciable elongation 
    (or equivalently, population of 
    the quadrupole mode); this cannot be captured by the linear stability 
    analysis of Ref.\ \cite{sinha}. An alternative, but 
    probably equivalent picture, is the one proposed in Ref.\ 
    \cite{kraemer}, which still holds if the rotation is switched on 
    moderately rapidly. 
    Another possibility could be that the 
    vortex creation is dependent on the excitation of higher than 
    quadrupolar modes due to trap imperfections (which in the 
    simulations would then be due to the square 
    numerical grid). But if this were the case, we would not have 
    expected the fair quantitative 
    agreement between numerics and experiment that is after all visible in 
    Fig.\ \ref{fig:phasediag}. (The occupation of modes during vortex 
    nucleation has been studied numerically in Ref.\ 
    \cite{Kasamatsu2002}, 
    but without special attention to the difference between 
    frequencies below and above the quadrupole frequency.) 
    We therefore conclude that the cause for 
    vortex excitation in this regime is not to be found in factors
    extrinsic to the zero-temperature Gross-Pitaevskii equation.

    \section{Stirring a condensate with a localized potential}\label{sec:stir}
    In the experiment reported in Ref.\ \cite{Raman2001a}, 
angular momentum was 
    imparted to the condensate by stirring it 
    with a pair of narrow repulsive laser beams, placed opposite to 
    each other and rotating about the center at a 
    radius $r_{s}$ with an angular frequency $\Omega$, so that the linear 
    velocity is $v_{s}=\Omega r_{s}$. The vortex formation is 
    in this case governed by quite different physics compared to the 
    previous section. In fact, there exists a crossover: 
    in the limit of a pair of wide beams at large $r_{s}$, the potential 
    becomes again that of a rotating anisotropic trap and the analysis 
    of the previous section applies. Here we shall, however, study the 
    case of narrow beams at both large and small distances from the 
    center. We take the potential to be a harmonic-oscillator one 
    plus two symmetrically placed repulsive Gaussian ``sticks'' revolving 
    about the origin:
    \begin{eqnarray}
    V(x,y) = \frac12 m \omega^2 (x^{2}+y^{2})
	+ V_{0} \left(e^{-\frac{|\rr-\rr_{0}(t)|^{2}}{2\sigma^2}}
	+ e^{-\frac{|\rr+\rr_{0}(t)|^{2}}{2\sigma^2}}\right).
    \end{eqnarray}
    The stirrer coordinates are given by $\rr_0(t)=(x_0(t),y_0(t))$ 
    with
    \begin{eqnarray}
	x_{0}(t) = r_{s}\cos\Omega t,\nonumber\\
	y_{0}(t) = r_{s}\sin\Omega t,
    \end{eqnarray}
    and the stirrer width $\sigma$ is smaller than the cloud size $\rtf$ but 
    not smaller than the healing length $\xi$.
    We have chosen to consistently work with a pair of stirrers in order
to facilitate comparison with the experimental data of Ref.
\cite{Raman2001a}. We have, however, found no qualitative differences 
when comparing with test runs made with one stirrer.

    The physics of the vortex creation is the following
    \cite{Jackson1998}. When an object or a 
    repulsive potential is moved through the condensate, pairs 
    of vortices with unit positive and negative circulation are created 
    in the wake. Positively 
    oriented vortices will be created to the left of the stirrer 
    relative to its direction of movement, and negatively oriented 
    vortices, antivortices, to the right. (This can be realized by 
    considering how the fluid flows back around the stirrer to fill the 
    wake.) The vortices and antivortices follow the 
    local current which initially is directed at an oblique angle, away 
    from the stirrer and in the direction of movement. The antivortices 
    will quickly drift out of the system while the positive vortices 
    may enter into the cloud depending on whether they have enough 
    energy to do so (assuming that the stirrer rotates counterclockwise 
    around the center; otherwise the role of vortices and antivortices 
    is interchanged). 
    The whole process is turbulent and is 
    complicated by inter-vortex interaction, sound waves 
    and density variations. 

The conditions for vortex creation are thus twofold: the linear 
velocity of the stirrer must be large enough to create vortices, and 
the angular velocity of the stirrer must be large enough to contain 
the vortices in the interior of the cloud. We study first the former 
requirement.
Vortex pairs are only created when the stirrer velocity exceeds a 
critical value given by the Landau criterion, Eq.\ (\ref{landau}).
For the situation in Ref.\ \cite{Raman2001a}, the 
prediction of Ref.\ \cite{Anglin2001} is $v_c=0.24\,c$. 
However, this analysis was carried out for a stirrer that moves close 
to the edge of the condensate. For the case of 
a Gaussian stirrer that moves 
in the bulk of the condensate, the abrupt potential gradient is expected 
to result in a lower critical velocity $v_c$ than that given by 
the surface-mode calculation \cite{ourfootnote}. In this case, a critical 
velocity about an order of magnitude smaller than the sound velocity $c$
has been observed numerically \cite{Jackson2000} and 
experimentally \cite{Raman2001a,Onofrio2000}. An approach that inserts the 
calculated energy and momentum of a vortex pair into the Landau 
criterion, Eq.\ (\ref{landau}), has yielded a critical velocity 
of the correct magnitude \cite{crescimanno}.

We have numerically investigated the critical velocity for the case when 
the coupling parameter $Na/\aos=56$, giving a Thomas-Fermi radius 
$\rtf=5.5\,\aos$, and a stirrer moving at a radius $r_s=0.2\,\rtf$ from the 
center. When the stirrer size $\sigma=0.05\,\rtf$, the critical 
velocity $v_c\approx 0.45\,c$ 
and when $\sigma=0.1\,\rtf$, $v_c\approx 0.35\,c$. 
Due to 
the noisy character of the data, the error bars on the observed critical 
velocities are about $\pm 0.1c$. This critical velocity is larger than 
that of Ref.\ \cite{Raman2001a}, but the difference in coupling 
strength is quite drastic: due to numerical constraints we have 
chosen to operate with a coupling strength $Na/l_z$ that is smaller 
than that of Ref.\ \cite{Raman2001a} by approximately a factor 10.

When the stirrer velocity is well above the critical one so that the
number of vortices is large, the build-up of angular momentum
resulting from the complex process of vortex creation and entry 
can be described classically.
The torque exerted on the fluid by the moving disturbance is estimated 
as follows. Assuming that the stirring potential $V_0$ is strong enough, 
the fluid that is instantaneously right 
in front of it is accelerated to the velocity of the stirrer $v_{s}$. 
The momentum increase per unit time due to one stirrer is therefore 
equal to the product of the mass of the fluid occupying an area swept out
by the stirrer in  unit time and the increase in velocity of the fluid at
radius $r_{s}$:
\beq
\label{momentumincrease}
	\frac{dp}{dt} = 
	2\sigma v_{s} m n_{2D}(r_s) (v_{s}-v(r_{s})),
\enq
where $n_{2D}(r)$ is the two-dimensional density of the fluid; 
$n_{2D}(r) = n_{3D}(r) l_z$. The width of the stirrer is taken to
be $2\sigma$.  The momentum increase
is azimuthally directed, and therefore the torque acting on the gas is
$r_{s}$ times the quantity on the  right-hand side of Eq.\
(\ref{momentumincrease}). When two stirrers are used, the momentum
increase is doubled and the total torque is 
\beq\label{torque1}
	\tau = 2 \cdot 2\sigma v_{s} r_s m(v_{s}-v(r_{s})) n_{2D}(r_s).
\enq
The effect of the
ensuing turbulent behavior, including the creation of  vortex pairs, is
to quickly distribute the momentum over the  whole cloud, and
when the number of vortices is large, the fluid rotates on average
as a solid body: $v(r) = \omega r$.
We can assume that the redistribution
of momentum happens instantaneously, because the  characteristic
timescale for the spreading of the disturbances is
$t_{\mathrm{spread}}=\rtf/c$. 
For realistic parameter values, this
timescale is indeed much shorter than the timescale for the growth of
the angular momentum, and therefore the system can at all times be 
assumed to be in solid-body rotation. 
However, during the stirring, we do not expect the vortex array to
extend outside the radius $r_s$ of the stirrer. Outside the region
enclosed by the path of the stirrer, the fluid velocity is the sum
of velocity fields from the individual quantized vortices, decaying as
the inverse distance from the vortex. A reasonable assumption
for the instantaneous fluid velocity is therefore
\begin{eqnarray}\label{fluidvelocity}
v(r) = \left\{ \begin{array}{ll}
\omega(t) r, & r < r_s;\\
\frac{\omega(t) r_s^2}{r}, & r > r_s,
\end{array}
\right. 
\end{eqnarray}
where the instantaneous solid-body rotation frequency $\omega(t)$
characterizes the rotation rate. Multiplying the local velocity, 
Eq.\ (\ref{fluidvelocity}), with $r$ and the Thomas-Fermi density,
Eq.\ (\ref{tfprofile}), and integrating over the whole cloud, one obtains 
the total angular momentum per particle at time $t$, 
\beq\label{l1}
L(t)=m\omega(t) r_s^2F\left(\frac{r_s}{\rtf}\right), 
\enq
where the function
$F(x)=1-x^2+x^4/3$. The
derivative of the angular momentum is to be equated to the torque, Eq.\
(\ref{torque1}), and the substitution of the angular momentum $L$ for 
the fluid velocity $v(r_s)$ according to Eq.\ (\ref{l1}) 
results in
\begin{equation}
    \dot{L} = 2\frac{2\sigma v_s n_{2D}(r_s)}{F(r_s/\rtf)} 
   	\left(m r_sv_s F(r_s/\rtf)-L\right).
    \label{eq:acceleration}
\end{equation}
Solving Eq.\ (\ref{eq:acceleration}) for $L(t)$, one obtains the time
dependence
\begin{equation}
    L(t) = L_{\mathrm{fin}} \left(1-e^{-2R t}\right),
    \label{eq:rateequation}
\end{equation}
where the final angular velocity is
\beq
  L_{\mathrm{fin}} = m r_s v_s \left(1 - 
	\left(\frac{r_s}{\rtf}\right)^2 + 
	\frac13 \left(\frac{r_s}{\rtf}\right)^4
	\right),
\label{finall}
\enq
and the rate constant:
\beq
	R = \frac{2\sigma v_s n_{2D}(r_s)}{F(r_s/\rtf)}.
\label{rate}
\enq 
The factor 2 multiplying the rate constant in Eq.\ 
(\ref{eq:rateequation}) is again the factor that reflects the number of 
stirrers. The torque is given by
\beq
\label{torque}
  \tau=2\cdot 2m\sigma r_s v_{s}^2 n_{(2D)}(r_s) e^{-2R t}.
\enq
The maximum value of the final angular momentum
$L_{\mathrm{fin}}$ always occurs when 
$r_s/\rtf = \sqrt{(9-\sqrt{21})/10} \approx 0.66$. The fact that the 
angular momentum is optimized when the stirrer is halfway between the 
center and the edge of the cloud is simply due to the fact that  
the linear velocity of the stirrer has been held constant. The 
maximum angular velocity is thus determined by a balance between on 
the one hand the increase in the amount of fluid that is enclosed by the 
path of the stirrer when $r_s$ is increased, and on the other hand 
the decrease in the angular velocity $\Omega$. If $\Omega$ is held 
constant instead of $v_s$, then the angular momentum will indeed be 
maximized by putting the stirrer at the edge of the condensate, as 
can be seen by substituting $\Omega r_s$ for $v_s$ in Eq.\ (\ref{finall}) 
and differentiating.

The classical model is valid in the case when many vortices are 
created, so that the assumption of solid-body rotation holds. 
The predicted final angular momentum $L_{\mathrm{fin}}$, Eq.\ 
(\ref{finall}), should therefore be much larger than unity; since the 
stirrer radius $r_s$ is on the order of the Thomas-Fermi radius 
$\rtf$, this translates to the criterion
\beq
\frac{v_s}{\aos\omega} \gg \left(\frac{l_z}{Na}\right)^{1/4}.
\label{lcond}
\enq
The classical model is thus easily satisfied when the coupling is 
strong, as expected. 
In addition, the model is expected to break down when the stirrer is in 
the vicinity of the cloud radius, $r_s \approx \rtf$, because of the 
Thomas-Fermi approximation. The rate $R$ is also 
ill described by the classical model when it is small; requiring $R 
\gg t_{\mathrm{spread}}^{-1}$ (see below Eq.\ (\ref{torque1})), we 
obtain the condition
\beq
\left(\frac{\sigma}{\aos}\right)\left(\frac{v_s}{\aos\omega}\right)
\gg \frac{a}{l_z} \sqrt{\frac{l_z}{N a}}.
\label{rcond}
\enq

\begin{figure}
\includegraphics[width=\columnwidth]{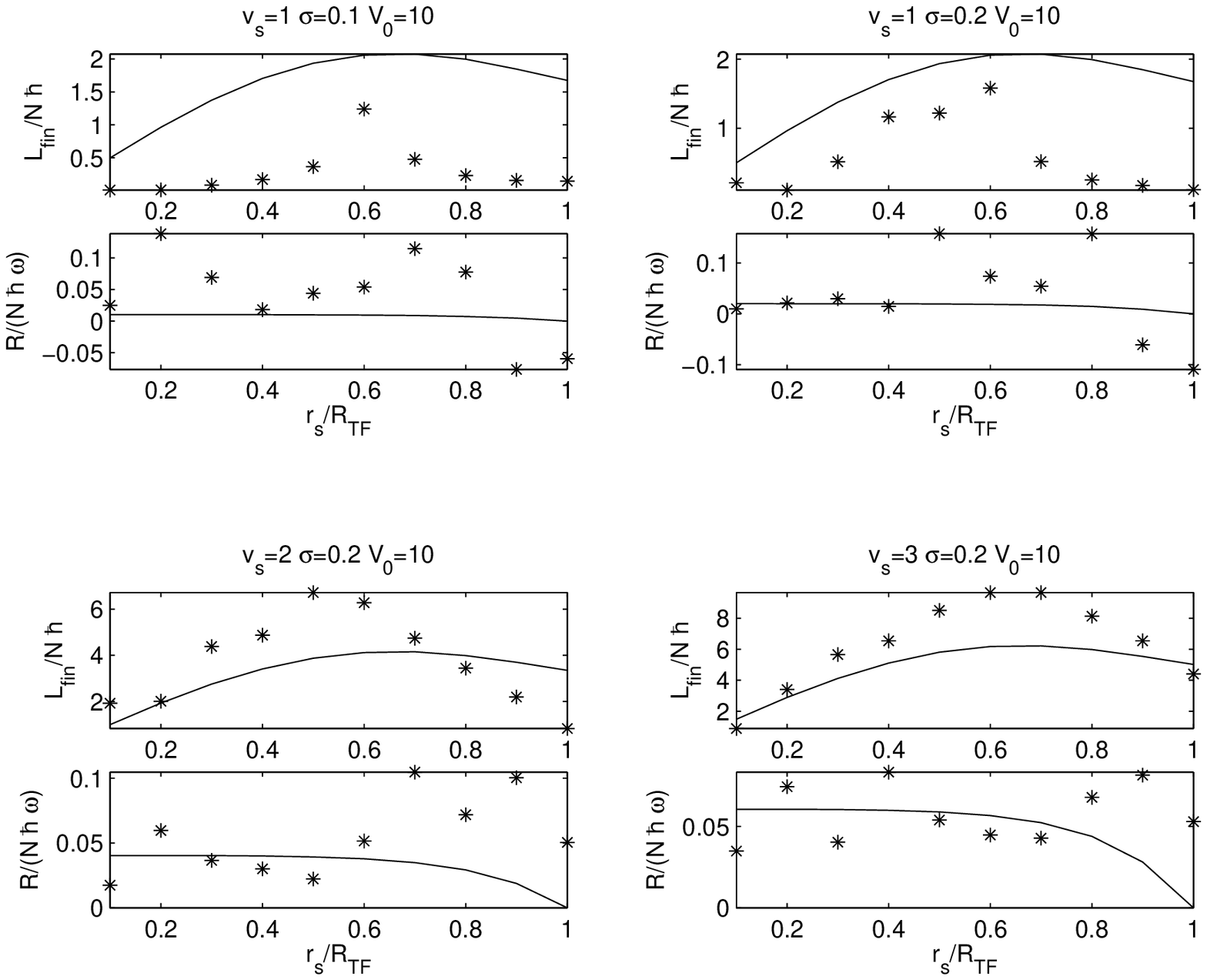}
\caption[]{Final angular momentum (upper panels) and initial rate of 
increase of angular momentum (lower panels) for a condensate stirred
by two rotating laser beams,
as functions of the distance of the stirrers from the trap center.
The stirrer velocity $v_s$, stirrer 
width $\sigma$ and peak stirrer potential $V_0$ are stated above each 
panel in units of the trap frequency $\omega$, oscillator length 
$\aos$, and trap energy $\hbar\omega$.
The lines are the classical equations explained in the text and the 
asterisks connected with lines are numerical results from the 
two-dimensional Gross-Pitaevskii equation. The coupling strength is chosen 
to $Na/l_z=40$, whereby the sound velocity $c\approx 3.6$ and 
Thomas-Fermi radius $\rtf\approx 5.0$ in trap units. The agreement with the 
classical prediction is best in the panels on the lower right, where the 
number of vortices is the largest.
\label{fig:radiussweep}}
\end{figure}

\begin{figure}
\includegraphics[width=\columnwidth]{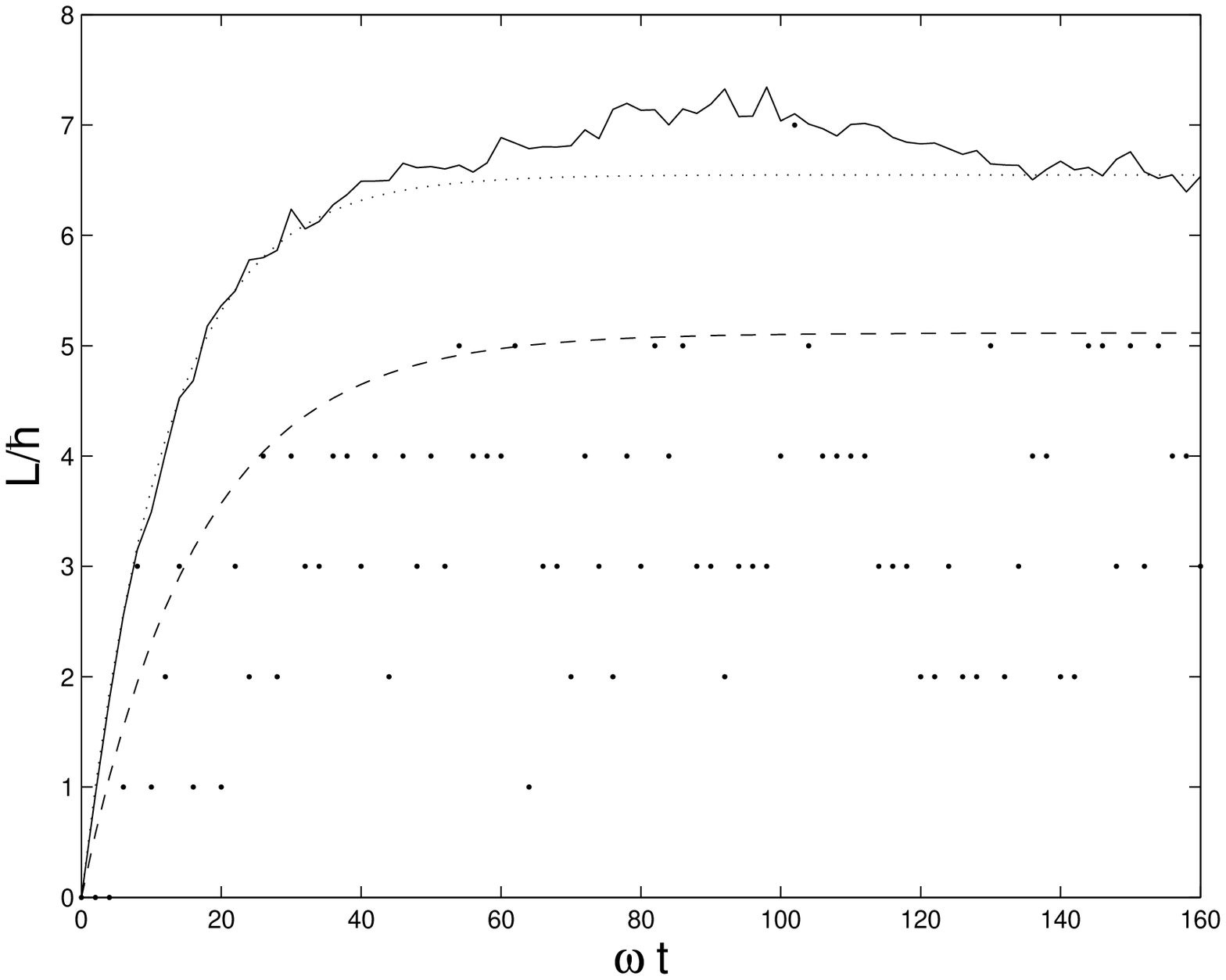}
\caption[]{Time evolution of the angular momentum and the number of 
  vortices of a condensate stirred by two rotating laser beams. The solid 
  line denotes the numerically computed angular momentum, the dashed line 
  is the classical prediction, the dotted line is a fit of the solid curve 
  to a function of the form (\ref{eq:rateequation}),
  and the dots signify the number of vortices 
  in the interior of the cloud. The parameter values are the same as in 
  the lower left two panels of Fig.\ \ref{fig:radiussweep}: 
  $v_s=3\omega\aos$, $\sigma=0.4\aos$, and we have chosen $r_s=0.6\rtf$.
\label{fig:l-stirrer}}
\end{figure}

The results of the numerical calculations are found to agree well 
with the classical model. Calculations have been performed on a 
two-dimensional system with $Na/l_z=40$, yielding a ratio 
between the Thomas-Fermi radius and the healing length 
$\rtf/\xi\approx 25$, so that the Thomas-Fermi approximation is 
applicable. The sound velocity is $c\approx 3.6\,\omega\aos$. The 
stirrer velocity has been varied between $\omega\aos$ and 
$3\omega\aos$.
Fig.\ \ref{fig:radiussweep} displays 
the final angular momentum $L_{\mathrm{fin}}$ of a 
two-dimensional cloud 
as a function of $r_{s}$ with fixed stirrer velocity $v_{s}$, as
well as the initial growth rate $R$ (found by fitting the data to
a function of the form of Eq.\ (\ref{eq:rateequation})). 
An 
illustrative example of the variation
of angular momentum with time for a particular value of $r_s$ is
displayed in Figure \ref{fig:l-stirrer}. We have chosen the parameters so 
that the agreement is fairly good, but not perfect. However, the 
qualitative appearance of the $L$ curve is the same for all parameter values.
The data is noisy because of the turbulent process, 
especially for the rate constant which is difficult
to determine accurately,
but on average it is seen to agree fairly well with Eqs.\
(\ref{eq:rateequation}-\ref{rate}) for higher values of $v_s$ and 
$\sigma$ so that the conditions 
(\ref{lcond}-\ref{rcond}) are met.

\begin{figure}
\includegraphics[width=\columnwidth]{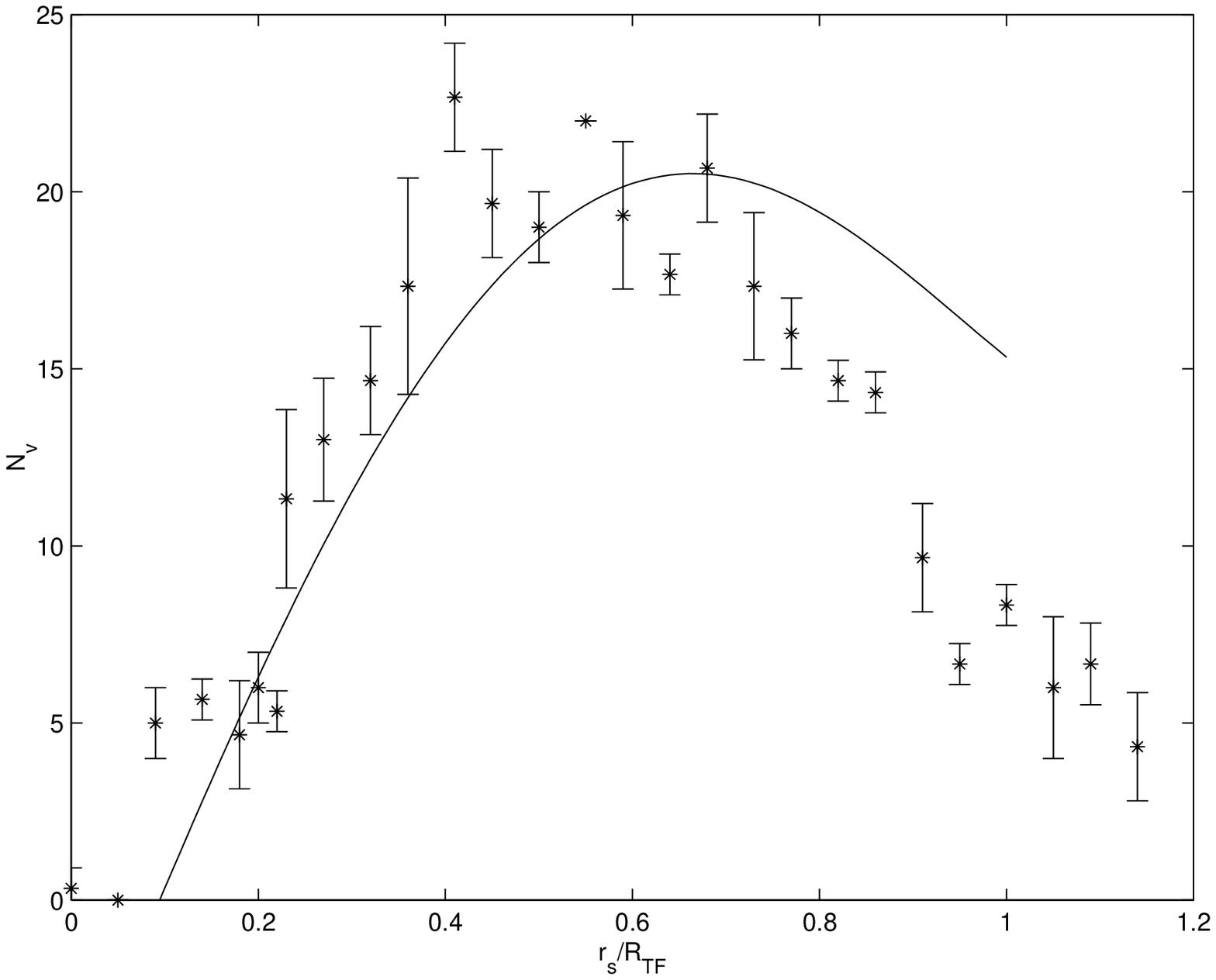}
\caption[]{Number of vortices after stirring as a function of 
the stirrer position for the conditions of the MIT experiment of 
Ref.\ \cite{Raman2001a}. 
The solid line depicts the classical model and the asterisks with 
error bars are the experimental results.
\label{fig:mit}}
\end{figure}

The classical model is seen to describe well the experiment reported in 
Ref.\ \cite{Raman2001a}, even though that experiment was performed in 
a prolate trap with anisotropy $\omega/\omega_z=4.3$.
For this trap, the Thomas-Fermi radius in the radial direction is 
$\rtf=12.6\aos$, and the speed of sound in the center of the trap is
$c=8.5\omega\aos$, where $\omega$ and $\aos$ are the radial trap 
frequency and oscillator length, respectively.
With a linear stirrer 
velocity $v_s = 0.2\,c$, the number of vortices $N_v$ 
was recorded as a function of the stirrer radius $r_s$. In the limit 
of a large vortex density, the fluid is on the average in solid-body 
rotation, for which $L= \hbar N_v/3$. However, we are here operating 
with a relatively small number of vortices, the maximum being about 20 
vortices in the experiment. Corrections to the solid-body result may 
therefore be important. Indeed, by calculating the angular momentum for a 
triangular vortex lattice in a cylindrically symmetric Thomas-Fermi 
cloud for different values of the lattice constant, we find the corrected 
number of vortices as a function of angular momentum to be well fit by 
the linear relation 
\beq\label{nv}
N_v\approx 3\frac{L}{\hbar} - 6.
\enq
This formula obviously breaks down 
when the number of vortices is close to 1, but is accurate for 
$N_v \geq 10$.
Inserting the numerical values into Eq.\ (\ref{finall}) and using 
Eq.\ (\ref{nv}) for the number of vortices, we 
obtain for the specific conditions of Ref.\ \cite{Raman2001a} the 
relation 
\beq
N_v = 64 x\left(1-x^2 + \frac13 x^4\right) -6,
\enq
where $x=r_s/\rtf$. The peak value of this function (at $x=0.66$) is
approximately $20.5$. This result is displayed in Fig.\ 
\ref{fig:mit}, and it is seen that the agreement is good, despite the 
drastic difference in geometry (prolate and two-dimensional, 
respectively).

\section{Conclusions}\label{sec:conclusions}
The nucleation of vortices in Bose-Einstein condensed gases contained
in time-dependent traps has been 
studied numerically for two different kinds of geometry. In the case of a 
rotating, deformed harmonic potential, vortices are nucleated when the 
anisotropy of the trap exceeds a critical value. Above 
the critical frequency for excitation of quadrupole modes, the 
numerically calculated 
critical anisotropy is in accordance with previous theoretical and 
experimental results. Below the quadrupole frequency, the numerical 
results are again in accordance with previous experimental results, 
showing that the nucleation of vortices in this regime can indeed 
take place at zero temperature. 
For a cloud stirred by a narrow moving repulsive potential, a 
classical model has been found to describe accurately the process of 
angular-momentum buildup. 
The classical argument is based on the fact that the velocity 
pattern is approximately a solid-body rotation in the interior of the 
system, where the density of vortices is high, but potential outside 
the path of the stirrer. As a result, the angular momentum approaches 
its final value exponentially. This model is applicable when the 
number of vortices is large, i.~e.\ when the stirrer velocity and 
width are not too small.
    
\begin{acknowledgments}
The authors acknowledge stimulating discussions with David Feder. 
We are grateful to Eleanor Hodby and to Chandra Raman 
for sharing experimental data 
with us. We acknowledge the Academy of Finland (Project 50314) for 
financial support. In addition, 
E.~L.\ is supported by the European Network ``Cold Atoms and 
Ultra-precise Atomic Clocks''. 
J.-P.~M.\ is supported by the Stichting voor Fundamenteel Onderzoek der 
Materie (FOM), which is supported by the Nederlandse Organisatie voor 
Wetenschaplijk Onderzoek (NWO).

\end{acknowledgments}

\end{document}